Submitted to Future Generation Computer Systems Journal (Elsevier)

# Task Offloading and Resource Allocation for Deterministic Service Provisioning in the IoT–Edge–Cloud Continuum

K. Aghababaiyan*, B. Coll-Perales*, and J. Gozalvez*

*Networked Systems Laboratory, Universidad Miguel Hernández de Elche, 03202 Elche, Spain

*Abstract*—**Future networks will seamlessly integrate computing and communication resources within a unified ecosystem of IoT devices, edge nodes, and cloud servers, forming an IoT-edge-cloud continuum. Within such an ecosystem, tasks can be dynamically offloaded to leverage resources across the continuum and efficiently scale with demand. Task offloading has traditionally focused on minimizing task execution latency or energy consumption. In contrast, we propose a deterministic task offloading and resource allocation scheme for the joint management of communication and computing resources to support services with deterministic service-level requirements. The proposed scheme aims to maximize the number of tasks completed on time, rather than minimizing the latency of individual tasks. To do so, it exploits variable task latency requirements to distribute the workload across the continuum. It can also prioritize high-priority tasks under resource scarcity. Numerical evaluations demonstrate that a deterministic task offloading strategy scales more effectively and can better guarantee timely task completion by balancing resource utilization across the continuum.**

## I. Introduction

The Internet of Things (IoT) is driving transformative changes across various domains, including connected and automated mobility and smart manufacturing. Connected vehicles rely on their own sensors, communication systems, and processing units to exchange real-time data for safer and more efficient driving tasks [1]. In the industrial IoT (IIoT) domain, collaborative robots (cobots) and autonomous guided vehicles dynamically interact with their environment for a more efficient execution of manufacturing and logistics tasks [2]. Critical services require efficient data processing for timely decision-making, which can place significant processing demands on computing resources as the complexity and number of tasks grow and requires careful latency management to meet deadlines of time-sensitive operations. In this context, the IoT-edge-cloud continuum has emerged as a key enabler to address increasing computational and latency challenges. The IoT-edge-cloud continuum seamlessly integrates computing, connectivity, intelligence, and services within a unified ecosystem of IoT devices, edge nodes, and cloud servers, and offers the possibility to dynamically plan and offload tasks' execution across resources distributed in the continuum [3]. To this aim, it is necessary to jointly manage communication and computing resources across the continuum.

To date, a strong focus has been on joint management schemes that minimize the task execution time. However, in critical verticals such as connected and automated mobility and smart manufacturing, it is usually more important to satisfy stringent deterministic service constraints - particularly in terms of bounded deadline compliance [4] than minimizing latency. Failure to meet these constraints may result in unsafe situations, operational disruptions, or system instability. The capacity to meet these constraints is

The work of K. Aghababaiyan is supported in part by the European Union under the 2024 MSCA Postdoctoral Fellowship program (project no. 101207598). This work has been also partially funded by the European Commission Horizon Europe SNS JU 6G-SHINE (GA 101095738) project, and by MCIN/AEI/10.13039/501100011033 (PID2020-115576RB-I00, PID2023-150308OB-I00), UMH *and Generalitat Valenciana* (CIAICO/2024/167).
*(Corresponding author: Keyvan Aghababaiyan).*

Email address: {kaghababaiyan, bcoll, j.gozalvez}@umh.es

strongly conditioned by the resources available in the continuum, the context conditions, and the service requirements. Connected vehicles typically feature considerable onboard processing capabilities but operate in highly dynamic environments with fluctuating connectivity conditions and diverse service requirements. In addition, software-defined vehicles need to support functionalities updated and expanded over time with the same computing resources. Smart factories usually operate within (semi-) static and controlled network infrastructures but deploy a large variety of devices and require collaborative efforts between devices for manufacturing task execution. In this context, achieving the appropriate balance between local processing and task offloading is essential. While offloading to edge or cloud servers can expand available computing capacity, it may introduce additional latency. On the other hand, relying solely on local execution may be limited by the local processing capabilities. The key challenge lies in designing task offloading strategies that holistically manage resources across the continuum in a manner that prevents communication and computing bottlenecks, preserves balanced resource utilization, and guarantees deterministic service levels [5].

Most existing latency-based task offloading schemes [6]-[9] focus on minimizing individual task latency, which risks overloading resources if too many tasks are assigned to the nodes that individually can minimize latency, causing the system to operate in a locally optimal but globally inefficient manner. This paper advances the state of the art with a deterministic task offloading and resource allocation strategy designed to satisfy deterministic service-level requirements while effectively scaling by balancing workloads across the continuum. The proposed scheme focuses on completing tasks within bounded deadlines rather than minimizing individual task latency. By exploiting the available task latency budget, it balances workload distribution across the IoT–edge–cloud continuum, which helps prevent computing and communication congestions in specific computing units and communication links. The scheme has the ability to prioritize the timely execution of critical tasks under resource scarcity. Its potential and adaptability to heterogeneous IoT operational settings are demonstrated in representative connected mobility and smart manufacturing scenarios characterized by different processing capabilities and diverse task requirements. Numerical evaluations show that the proposed deterministic strategy can scale better and successfully complete a higher number of tasks on time than state-of-the-art schemes focused on minimizing individual task latency thanks to a more balanced workload distribution across the continuum.

The remainder of this article is organized as follows. Section II reviews the state-of-the-art in task offloading and resource allocation. Section III describes the system model adopted for task offloading and resource allocation across the IoT-edge-cloud continuum. Section IV presents the proposed deterministic task offloading and resource allocation scheme. Section V outlines the simulation setup, evaluation parameters and scenarios as well as the benchmark schemes. Section VI analyzes the evaluation results, and Section VII summarizes the key findings.

## II. State of The Art

A key aspect in managing the IoT-edge-cloud continuum is determining the optimal distribution of task processing across resources in the continuum. This should be done noting that not all IoT nodes are resource-constrained; for instance, devices such as Connected and Automated Vehicles (CAVs) often feature powerful local processors. In recent years, growing attention has been devoted to developing task offloading and resource allocation schemes across different computing nodes of the continuum - ranging from local-edge cooperation to cloud-based and fully integrated IoT-edge-cloud continuum - with the goal of minimizing latency and improving the efficiency of distributed computing. The work in [10] focuses on latency-aware task execution in mobile edge computing, where computation occurs locally or at nearby edge servers. It optimizes resource allocation and offloading decisions to reduce latency, though it remains confined to the edge and local IoT nodes. Studies [6] and [7] expand this model by employing the computing capacity of interconnected IoT nodes, specifically vehicles, as distributed resources. In [6], parked and moving vehicles cooperate with an edge server to process offloaded tasks and minimize total priority-weighted processing latency through coordinated offloading among vehicles, and between vehicles and the edge. The study reported in [7] introduces greater flexibility by enabling vehicles to decide whether to execute tasks locally, offload them to a roadside unit, or share them directly with other vehicles. These studies emphasize the role of vehicles as mobile IoT devices that augment edge resources, contributing to latency reduction. [8] introduces an inverse offloading framework, where tasks are transferred from the edge back to vehicles. This scheme alleviates the computational burden on the edge server while maintaining low latency by effectively exploiting underused vehicular computing capacity. [11] focuses on cloud-centered task migration, and designs strategies that move computation from onboard vehicle systems to cloud servers to support cooperative processing. The study examines how centralized execution and 5G connectivity can meet increasing computational demands while maintaining latency below the required deadline constraints. [9] and [12] advance the concept toward multi-layer collaboration, forming a full IoT–edge–cloud continuum with the goal of minimizing latency. In [9], computation is organized across the full continuum, distinguishing between latency-sensitive and non–time-sensitive tasks, with the objective of minimizing latency for latency-sensitive tasks. [12] integrates idle IoT and vehicular devices as intermediate computing nodes that cooperate with edge and cloud resources to minimize latency. Studies [13]-[15] focus on load balancing and spatial load distribution across resources in the continuum. In [13] and [14], the authors introduce schemes that identify overloaded edge nodes and reassign tasks or users to underutilized ones, thereby maintaining responsiveness and preventing congestion. [15] proposes a location-aware allocation scheme that redistributes computational tasks from congested regions to less utilized areas, improving overall system stability and latency consistency.

Collectively, these studies highlight the potential of the IoT–edge–cloud continuum, and the importance of task offloading schemes to effectively exploit the continuum. To date, contributions have mostly focused on minimizing individual task execution latency. However, this focus increases the risk of overloading certain computing units and communication resources, while many vertical services require deterministic service levels and not necessarily minimizing the latency of each task. This paper addresses this gap with a deterministic task offloading and resource allocation strategy that prioritizes the timely completion of tasks within predefined deadlines and a balanced distribution of workloads across the continuum for better scaling. An initial version of this work was presented in [16]. This manuscript provides a robust extension to that past study by introducing a load distribution mechanism and a priority-based version of the proposed scheme. In addition, we conduct a thorough analysis on the impact of distinct operational environments—such as connected vehicular mobility and smart industrial manufacturing scenarios—demonstrating the performance, robustness, and adaptation of the proposed strategy under diverse network and device characteristics.

## III. System Model

We consider an IoT-edge-cloud continuum system composed of heterogeneous IoT devices connected to edge computing nodes and a cloud server through a wide-area wireless network. The system model is designed to capture a broad range of operational environments where tasks generated by IoT devices can be processed locally or offloaded to edge or cloud, subject to communication and computing constraints and task-specific latency requirements. To reflect the diversity of real-world deployments, the model encompasses two representative emerging IoT–edge–cloud continuum scenarios: connected mobility and smart manufacturing. These scenarios exhibit fundamentally different characteristics in terms of device heterogeneity, local processing capabilities, task generation patterns, and reliance on network-assisted computation.

The connected mobility scenario includes $J_{v_1}$ low-power wearable devices (e.g. smartwatches), $J_{v_2}$ general-purpose mobile devices (e.g. smartphones), and $J_{v_3}$ vehicles equipped with powerful onboard computing platforms where $J_v$=$J_{v_1}$+$J_{v_2}$+$J_{v_3}$. Tasks are generated by these three types of devices with proportions $\alpha_{v_1}, \alpha_{v_2}, \alpha_{v_3}$, where $\alpha_{v_1} + \alpha_{v_2} + \alpha_{v_3} = 1$. The smart manufacturing scenario considers a mix of $J_{I_1}$ mobile IoT devices, $J_{I_2}$ assembly robots, $J_{I_3}$ advanced factory robots, and $J_{I_4}$ autonomous guided vehicles with more constrained local processing capabilities than vehicles in the connected mobility scenario. $J_I = J_{I_1} + J_{I_2} + J_{I_3} + J_{I_4}$ is the total number of IoT industrial devices. Tasks are generated by different types of devices with proportions $\alpha_{I_1}, \alpha_{I_2}, \alpha_{I_3}$, $\alpha_{I_4}$, where $\alpha_{I_1} + \alpha_{I_2} + \alpha_{I_3} + \alpha_{I_4} = 1$.

Tasks $f_i$ , $i \epsilon \{1, \dots, I\}$ generated by the devices in both scenarios can be represented by the tuple $\{c_i, s_i, s'_i, G_i, D_i\}$. $c_i$ is the task $f_i$'s computational demand and $s_i$ is its associated size. When a task is delegated to the edge units or cloud server, the processed result must be transmitted back to the original unit. The size of processed results of $f_i$ is defined as $s'_i = \rho \cdot s_i$, where $\rho < 1$. $G_i$ is $f_i$'s generation time, and $D_i$ its deadline by which all processing must be completed. If the task is offloaded to the edge or cloud, the deadline accounts for the time required to transmit the task to the designated processing unit, the processing time, and the time needed to send the processed result back.

Local IoT ($j \epsilon J$; either $J_v$ or $J_I$), edge ($q \epsilon Q$) and cloud ($C$) computing units are modeled considering their processing power $\chi_m$ and maximum processing capacity $X_m$ over a time period $T$ with $X_m = \chi_m . T$, $m \epsilon M$; $M = (J, Q, C)$. The computing time for a task $f_i$ processed by computing unit $m$ is:

$$t_p^i = \frac{c_i}{\chi_m}. \quad (1)$$

We consider that the wireless link to connect the IoT devices to the wide-area network is based on a Frequency Division Duplexing (FDD)-based Orthogonal Frequency Division Multiple Access (OFDMA) radio access interface. The amount of information that can be transmitted by a resource block $\kappa \in K$ can be expressed as follows:

$$R_\kappa = T_k . BW_\kappa \times \log_2(1 + \gamma_\kappa)(1 - e_\kappa), \quad (2)$$

where $T_k$ is the duration of the resource, $BW_\kappa$ is the resource's bandwidth, $\gamma_\kappa$ is the Signal-to-Interference-plus-Noise Ratio (SINR), and $e_\kappa$ is the bit error rate, which is a function of the modulation and coding scheme (MCS) used. $K$ is the total number of orthogonal resource blocks, which are divided into $K^{UL}$ and $K^{DL}$ resources for the uplink and downlink bands. We assign each communication resource $\kappa$ an MCS determined by its SINR $\gamma_\kappa$. Channel fading is modeled with a Rayleigh distribution. Considering $K_i^{UL}$ as the total number of resource blocks used in the uplink for the offloading of the tasks $f_i$, the uplink channel data capacity can be estimated as:

$$R_i^{UL} = \sum_{\kappa \in K_i^{UL}} R_\kappa. \quad (3)$$

Similarly, the downlink channel data capacity $R_i^{DL}$ is computed as the sum of the capacities of the $K_i^{DL}$ resource blocks allocated in the downlink. It should be noted that $s_i \le R_i^{UL}$ and $s'_i \le R_i^{DL}$ must hold to ensure that the uplink and downlink communication links can successfully transfer the task $f_i$ and its processing result, respectively.

If task $f_i$ with size $s_i$ needs to be offloaded, its uplink transmission time is computed as:

$$t_{UL}^i = \frac{s_i}{R_i^{UL}}, \quad (4)$$

and the downlink transmission time for the processed result of this task can be expressed as:

$$t_{DL}^i = \frac{s'_i}{R_i^{DL}}. \quad (5)$$

When a task is offloaded to the edge or cloud, the time $T_i$ necessary to execute a task $f_i$ includes the processing time at the corresponding unit ($t_p^i$), the time to upload the task from the IoT device to the computing unit ($t_{UL}^i$), and the time to download the processed result back to the IoT device ($t_{DL}^i$):

$$T_i = t_p^i + t_{UL}^i + t_{DL}^i. \quad (6)$$

If the task is locally executed at the IoT device then $T_i = t_p^i$. A task is satisfactorily served if the total execution time $T_i$ is less than or equal to the task's deadline $D_i$.

## IV. Deterministic Task Offloading And Resource Allocation

This section presents the proposed *Deterministic* task offloading and resource allocation scheme, operating over the IoT–edge–cloud continuum system described in Section III. The proposed scheme manages the heterogenous communication and computing resources available across the continuum to satisfy tasks' bounded latency deadlines, rather than minimizing individual task execution time. To this aim, we define the penalty cost $H_D$ as

$$H_D = \sum_{i \in I} \Lambda(\tau_i), \quad (7)$$

and $\Lambda(\tau_i)$ is a penalty function defined as:

$$\Lambda(\tau_i) = \begin{cases} 0, & 0 \le \tau_i \le 1, \\ \Pi, & \tau_i > 1, \end{cases} \quad (8)$$

with $\tau_i = \frac{T_i}{D_i}$. This penalty function assigns a high cost $\Pi$ to tasks $f_i, i \in \{1, \dots, I\}$ whose execution time $T_i$ exceeds their deadline $D_i$, while tasks completed within their deadlines incur no penalty.

We initially formulate the task offloading and resource allocation problem as an optimization problem where the objective is minimizing the overall penalty cost $H_D$:

$$Min\ H_D \quad (9)$$

$$s.t. \quad C_1: a_{i,j} \in \{0, 1\}, \forall j \in J, \forall i \in I,$$

$$C_2: a_{i,q} \in \{0, 1\}, \forall q \in Q, \forall i \in I,$$

$$C_3: a_{i,C} \in \{0, 1\}, \forall i \in I,$$

$$C_4: \sum_{j \in J} a_{i,j} + \sum_{q \in Q} a_{i,q} + a_{i,c} \le 1, \quad \forall i \in I,$$

$$C_5: \sum_{i \in I} c_i a_{i,m} \le X_m, \quad \forall m \in M,$$

$$C_6: b_{i,\kappa}^{UL}, b_{i,\kappa}^{DL} \in \{0, 1\}, \forall i \in I, \forall \kappa \in \mathrm{K},$$

$$C_7: \sum_{i \in I} b_{i,\kappa}^{UL} = 1, \forall \kappa \in \mathrm{K}^{UL},$$

$$C_8: \sum_{i \in I} b_{i,\kappa}^{DL} = 1, \forall \kappa \in \mathrm{K}^{DL},$$

$$C_9: s_i \le \sum_{\kappa \in K_i^{UL}} R_\kappa, \forall i \in \mathrm{I},$$

$$C_{10}: s'_i \le \sum_{\kappa \in K_i^{DL}} R_\kappa, \forall i \in \mathrm{I},$$

$$C_{11}: b_{i,\kappa}^{UL} = 0, if\ t(\kappa) < G_i,$$

$C_{12}$: $\sum_{j\epsilon J} a_{i,j} + \sum_{q\epsilon Q} a_{i,q} + a_{i,c} = 0, if\ t < G_i + t_{UL}^i$,
$C_{13}$: $b_{i,\kappa}^{DL} = 0, if\ t(\kappa) < G_i + t_p^i + t_{UL}^i$.

Constraints $C_1$-$C_4$ use binary variables $a_{i,j}$, $a_{i,q}$ and $a_{i,c}$ to enforce that each task is allocated to a single computing unit, either in the IoT ($C_1$), edge ($C_2$), or cloud ($C_3$) domain. Constraint $C_5$ ensures that the selected computing unit is not overloaded by the computation demand of the allocated task. Constraint $C_6$ defines the allocation of communication resources to tasks in the uplink and downlink directions. Binary variables $b_{i,\kappa}^{UL}$ and $b_{i,\kappa}^{DL}$ indicate whether communication resource $\kappa\epsilon K$ is assigned to task $f_i$ - for offloading to the edge or cloud - for transmitting the processed result back to the IoT device, respectively. In line with OFDMA operation, constraints $C_7$ and $C_8$ ensure that each communication resource block is allocated to at most one task. Constraints $C_9$ and $C_{10}$ guarantee that the aggregated capacity of the allocated communication resources is sufficient to transmit the task in the uplink and its processed result in the downlink, respectively. Constraints $C_{11} - C_{13}$ enforcethe temporal ordering of operations: uplink transmission can only occur after task generation, task processing after the uplink ,transmission is completed and the downlink transmission takes place after task processing.

We extend the initial formulation in (7) to promote a balanced utilization of resources across the IoT-edge-cloud continuum. This extension is aimed at mitigating load concentration and resource bottlenecks under high traffic conditions, while preserving the *Deterministic* operation of the scheme. In this case, the objective function is still minimizing the overall penalty cost, but this cost is now expressed as:

$$H_{DD} = \sum_{i\epsilon I} \Lambda(\tau_i) + \lambda_1 \sum_{m\epsilon M} N_m . \left|\bar{\beta} - \beta_m\right| + \lambda_2 \sum_{m\epsilon M} (1 - \eta_m), \quad (10)$$

where the first term corresponds to the penalty function $H_D$ in (7) and accounts for tasks whose execution exceeds their deadline. The second term in (10) promotes a more balanced utilization of computing resources across the continuum by penalizing deviations from the continuum's average usage ratio. $N_m$ denotes the set of tasks assigned to computing unit $m$, and $\beta_m$ represents the usage ratio of computing unit $m$:

$$\beta_m = \frac{\sum_{i\epsilon I} c_i a_{i,m}}{X_m}, \quad (11)$$

where $\bar{\beta}$ is the average usage ratio across different computing units, which is computed as:

$$\bar{\beta} = \frac{1}{M} \sum_{m\epsilon M} \beta_m. \quad (12)$$

By weighting the deviation $\left|\bar{\beta} - \beta_m\right|$ with $N_m$, this second term in (10) places greater emphasis on load imbalance affecting computing units that serve a larger number of tasks. The second term in (10) discourages load imbalance and excessive task concentration in certain computing units. Finally, the third term in (10) penalizes low task satisfaction at each computing unit by weighting the fraction of tasks whose execution time $T_i$ exceeds their deadline. This fraction is equal to 1 - $\eta_m$, where $\eta_m$ is defined as the task satisfaction ratio for computing unit $m$. The satisfaction ratio is defined as the ratio of tasks completed within their deadlines. The third term is hence defined to encourage high per-unit satisfaction ratios and prevent the concentration of poorly served tasks on individual computing units. The second and third terms in (10) use $\lambda_1$ and $\lambda_2$ as weighting factors to control the importance of the terms. The objective function for the task offloading and resource allocation problem is then defined as:

$$Min \;\; H_{DD} \quad (13)$$
$$s.t. \qquad C_1 - C_{13},$$

where $C_1 - C_{13}$ are the constraints defined in (9). The objective function is further extended to incorporate task prioritization so that high-priority tasks can still be completed within their deadline under resource-constrained conditions. In this case, the penalty cost is defined as follows:

$$H_{PDD} = \sum_{i\epsilon I} \Lambda_P(\tau_i) + \lambda_1^H \sum_{m\epsilon M} N_m^H . \left|\bar{\beta} - \beta_m\right| + \lambda_2^H \sum_{m\epsilon M} (1 - \eta_m^H), \quad (14)$$

where $\Lambda_P(\tau_i)$ is defined as:

$$\Lambda_P(\tau_i) = \begin{cases} 0, & 0 \le \tau_i \le 1, \\ \Pi_{P_i}, & \tau_i > 1, \end{cases} \quad (15)$$

where $P_i \epsilon \{L, H\}$ denotes whether task $f_i$ is of low (L) or high (H) priority. The penalty values satisfy $\Pi_L \ll \Pi_H$, thereby enforcing a stronger penalty for deadline violations of high-priority tasks and biasing the optimization toward their timely completion under resource-constrained conditions. The second and third terms in (14) extend (10) by focusing specifically on high-priority tasks. To this end, in (14), $N_m^H$ denotes the number of high-priority tasks assigned to computing unit $m$, while $\eta_m^H$ represents the satisfaction ratio of high-priority tasks at unit $m$. By weighting the load imbalance with $N_m^H$, the second term discourages concentration of high-priority tasks on individual computing units and promotes a balanced distribution of critical workloads across the IoT–edge–cloud continuum. The parameters $\lambda_1^H$, $\lambda_2^H$ are weighting factors controlling the relative importance of load

**Algorithm I** : Genetic algorithm

**1. For** gen = 1 to $G$ do
2. Initialize population $\{a_{i,j}, a_{i,q}, a_{i,c}\}$ satisfying $C_1 - C_5, C_{12}$.
3. For each individual:
4. Allocate $b_{i,\kappa}^{UL}$ under $C_7, C_9, C_{11} \longrightarrow$ compute $t_{UL}^{i}$.
5. Calculate $t_p^i$ using (1).
6. Allocate $b_{i,\kappa}^{DL}$ under $C_8, C_{10}, C_{13} \longrightarrow$ compute $t_{DL}^{i}$.
7. Evaluate objective functions of (9), (13), (16).
8. Select top 25% as elite.
9. Generate 75% via crossover (elite).
10. Mutate 20% of new individuals.
**11. End For**
**12.** Return best $\{a_{i,j}, a_{i,q}, a_{i,c}\}$ across G generations balancing and maximizing the number of high-priority tasks completed by their deadlines per computing unit. The final objective function for the joint task offloading and resource allocation problem is defined subject to constraints $C_1 - C_{13}$ in (9) as:

$$Min \quad H_{PDD} \tag{16}$$
$$s.t. \qquad C_1 - C_{13}.$$

The optimization problem in (16) is NP-hard, and we use the genetic algorithm in Algorithm I to obtain near-optimal solutions [16]. The algorithm runs for $G$ =10 generations (line 1) with a population size of 1,000 individuals per generation. This configuration is sufficient because the large population size allows for broad exploration of the search space, leading to rapid convergence. By the tenth generation, the variation in the objective values stabilizes to less than 1%, indicating that further iterations would yield negligible improvements. Algorithm I initializes the population based on constraints $C_1 - C_5$ and $C_{12}$ (line 2). The objective functions are evaluated in lines 3 to 7. For each generation, 25% of the top-performing (elite) individuals are selected for crossover to form the next generation (lines 8 and 9). Additionally, in line 10, mutation is introduced in 20% of the population to enhance solution diversity and improve convergence toward a near-optimal solution [18].

## V. Evaluation Scenario

We consider two representative IoT-edge-cloud scenarios from the connected mobility and smart manufacturing verticals to evaluate the ability of the task offloading schemes to adapt to heterogeneous operational environments. These scenarios exhibit fundamentally different characteristics in terms of device heterogeneity, local processing capabilities, and reliance on network-assisted computation. The connected mobility scenario includes $J_{v_1}$=4 smartwatches, $J_{v_2}$=4 smartphones, and $J_{v_3}$=2 vehicles, resulting in a total of $J_v$=10 IoT devices. Tasks are generated by these devices at proportions of $\alpha_{v_1}$ =0.1, $\alpha_{v_2}$=0.5 and $\alpha_{v_3}$ =0.4, respectively. These values align with the standard distribution of tasks commonly observed across these specific device categories [19]. The computational workload of each task is modeled as a uniformly distributed random variable whose range depends on the originating device type. Tasks generated by smart vehicles are more computationally demanding, ranging from 75 to 150 Mcycles. In contrast, tasks generated by smartwatches and smartphones range from 25 to 35 Mcycles and from 30 to 60 Mcycles, respectively, reflecting the less demanding applications they support [20]. Following the specifications of commercial off-the-shelf products, the local processing capabilities are set to 400 MHz for smartwatches, 3 GHz for smartphones, and 12 GHz for vehicles, capturing the large disparity in onboard computing power across device types.

In the smart manufacturing scenario, we consider a medium-size industrial deployment [21] with $J_{I_1}$= 5 mobile IoT devices, $J_{I_2}$= 6 assembly robots, $J_{I_3}$ = 6 advanced factory robots, and $J_{I4}$= 6 automated guided vehicles, for a total of $J_I$ = 23 IoT devices. The four device types generate tasks with proportions of $\alpha_{I_1}$ =0.1, $\alpha_{I_2}$ =0.3, $\alpha_{I_3}$ =0.3, and $\alpha_{I_4}$ =0.3, respectively, reflecting the heavier workloads produced by robotic systems compared to simple sensing devices [22]. Their workloads are modeled as uniformly distributed random variables with ranges depending on the device type and their processing needs ([23], [24]): (25–35) Mcycles for mobile IoT devices, (30–100) Mcycles for assembly robots and advanced factory robots, and (60–120) Mcycles for autonomous guided robots. Their local processing power is set to 400 MHz for mobile IoT devices, 1.2 GHz for assembly robots, 2 GHz for advanced factory robots, and 2.5 GHz for autonomous guided vehicles [25].

In both scenarios, the infrastructure includes one edge computing node and one cloud server, with processing capacities of 70 GHz and 150 GHz, respectively [26].

Different workload scenarios are analyzed with the number of parallel tasks ranging from 35 to 80 every 100 ms, and tasks generated uniformly over time. The task deadlines are uniformly distributed between 50-100 ms for the connected mobility scenario, and 50-90 ms for the smart manufacturing scenario [27]. Representing typical payloads for sensing and control data [20], task sizes are uniformly distributed in the range of 1–2 Mbits, and the size of processed results is set to $\rho$=0.15 of the original task size. We consider two task priority classes. The penalty values $\Pi_{P_H}$ and $\Pi_{P_L}$ in (15) for high- and low-priority tasks are set to 100 and 1, respectively. The regularization factors in (10) and (14) are $\lambda_1 = \lambda_1^H = 2$ and $\lambda_2 = \lambda_2^H = 2$, respectively. These values are chosen to give equal importance to satisfying latency deadlines in each computing unit and ensuring balanced workload distribution across the continuum.

The FDD-based OFDMA wide area network is configured with a 100 MHz bandwidth, 30 kHz subcarrier spacing (SCS) and a time slot duration of 0.5 ms, typically used in 5G NR environments representing aggregated FDD carriers. Diverse communication conditions are analyzed by varying the SINR between devices and the base station in the range of 15 – 30 dB.

## VI. PERFORMANCE EVALUATION

This section compares the performance of the deterministic task offloading and resource allocation proposal against three benchmark schemes. With the first one (*Local*), all tasks generated by IoT devices are processed by the devices themselves. This scheme represents scenarios where task offloading is not feasible - e.g. when devices are not connected to the wide-area network - and local processing is the only option. The second scheme (*Offload*) offloads all tasks generated by the IoT devices for processing at edge nodes or cloud servers. This scheme represents scenarios where local computing resources are insufficient, and tasks must rely entirely on network-assisted processing. The third benchmark scheme (*Minimum*) follows the principles of [6]-[9] and jointly manages communication and computing resources with the objective of minimizing the execution time $T_i$ of individual tasks $f_i$. The objective function is defined as:

$$H_M = \sum_{i \epsilon I} T_i , \tag{17}$$

and we apply the same constraints $C_1 - C_{13}$ in eq. (*9*) for a fair comparison with the *Deterministic* problem. The optimization problem is then defined as follows, and we apply the same genetic algorithm for its resolution:

$$Min \quad H_M \tag{18}$$

$$s.t. \qquad C_1 - C_{13}.$$

Figure 1 compares the satisfaction ratio achieved by the *Deterministic* and benchmark schemes under varying traffic loads and channel quality conditions in the connected mobility scenario. Unless otherwise stated, the *Deterministic* proposal is configured with the penalty value $\Pi$ in (8), assigned to tasks that are not completed by their deadlines equal to 100; this value is chosen to be higher than the maximum number of tasks to impose a sufficiently high penalty. The satisfaction ratio is defined as the ratio of tasks completed within their deadlines. Figure 1 shows that the *Deterministic*

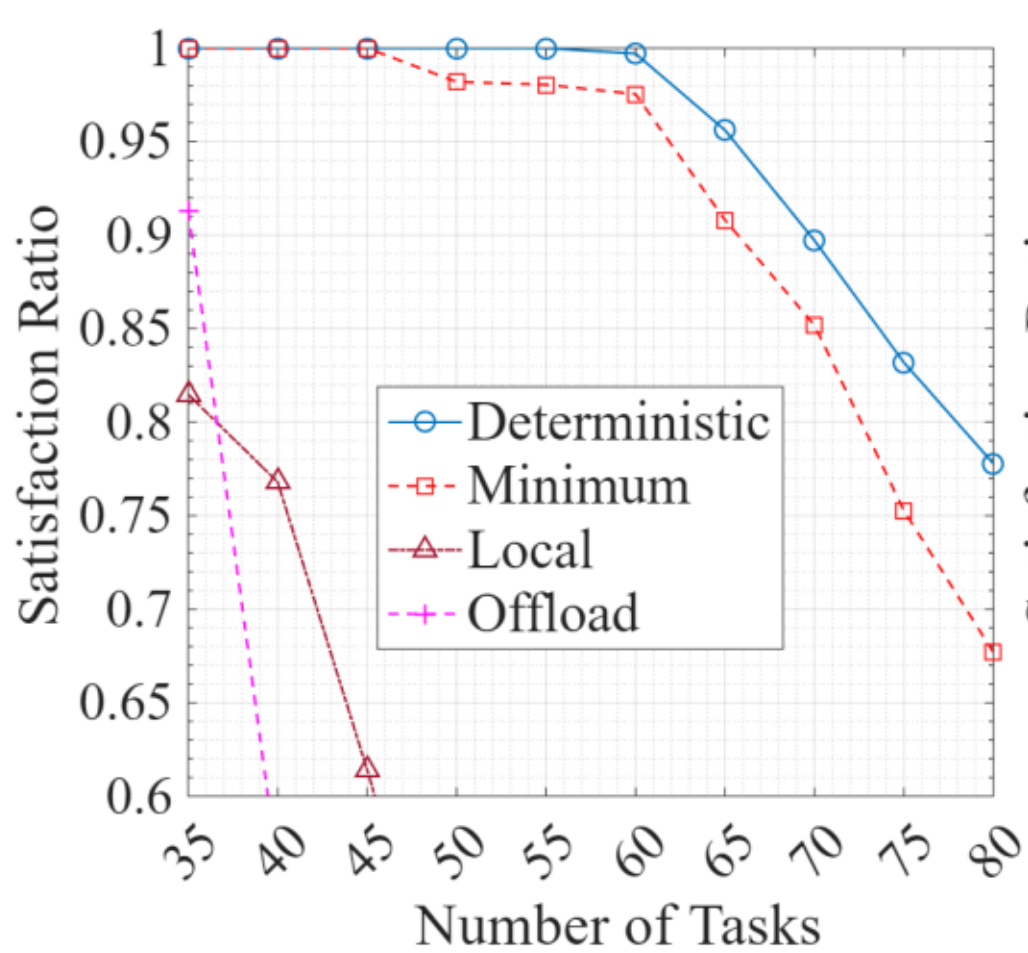

(a) Satisfaction ratio as a function of the number of tasks (SINR= 30 dB)

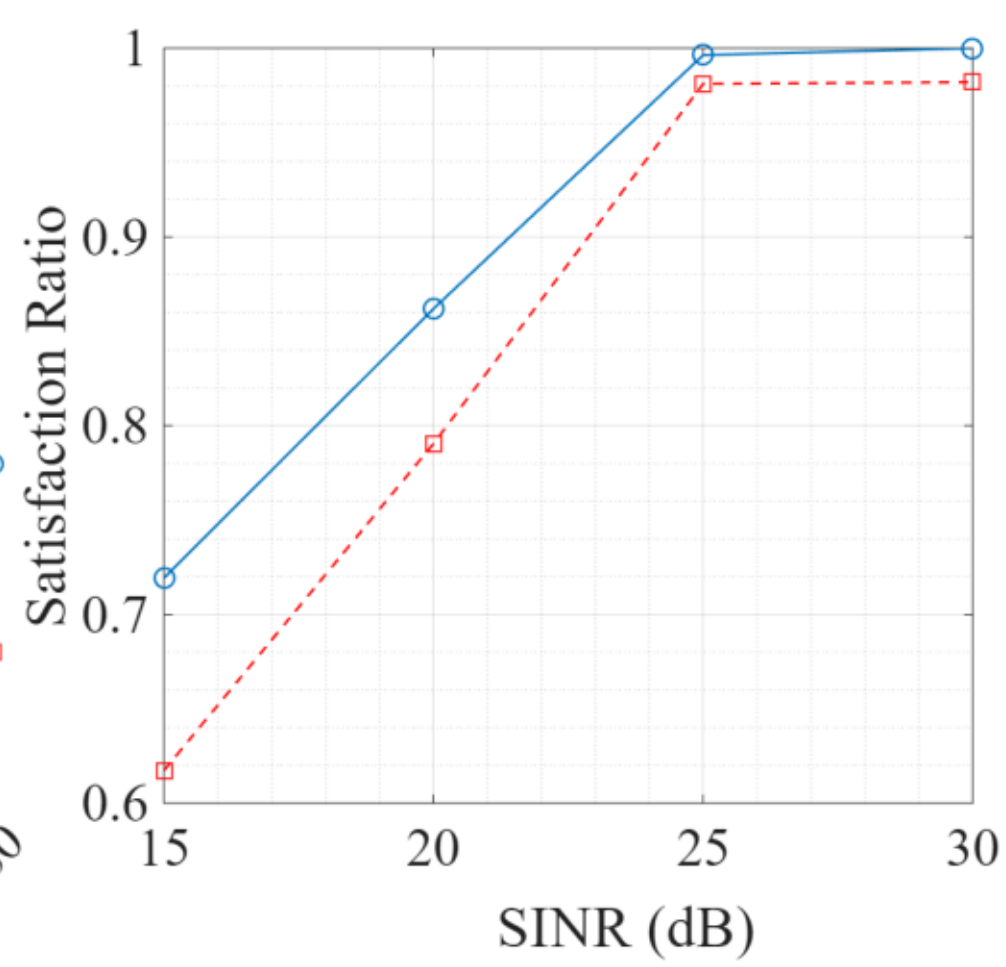

(b) Satisfaction ratio as a function of the SINR (number of tasks=50)

Figure 1. Satisfaction ratio in the connected mobility scenario.

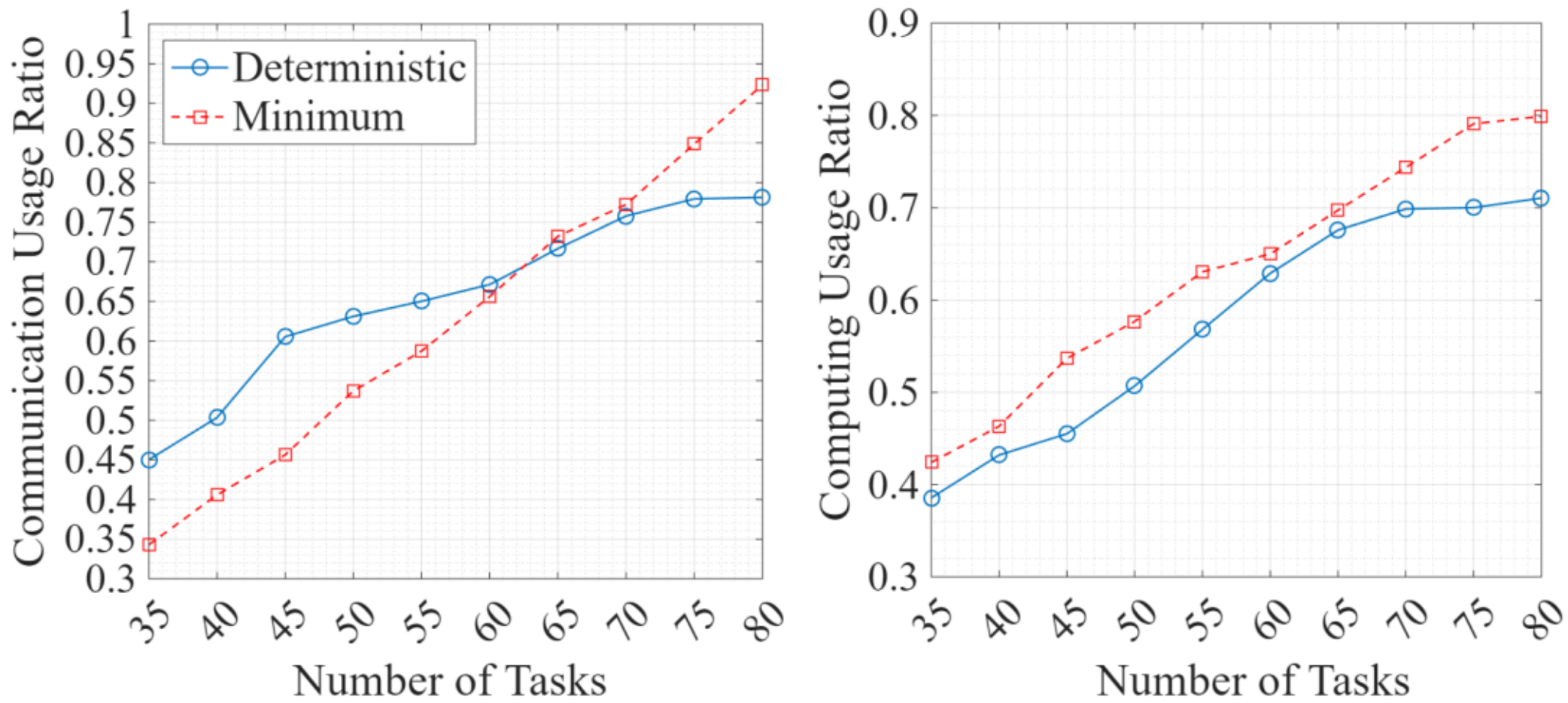

(a) Communication resources usage (b) Computing resources usage

Figure 2. Resources usage ratio as a function of number of tasks for SINR= 30 dB in the connected mobility scenario.

proposal achieves the highest satisfaction ratio across all considered traffic loads and channel quality conditions. This confirms the effectiveness of explicitly targeting deadline satisfaction when allocating communication and computing resources. Figure 1 also highlights the benefits of exploiting an integrated IoT-edge-cloud continuum with the *Deterministic* and *Minimum* schemes compared to Local and Offload. The latter experience a pronounced degradation as the number of tasks increases, since they are restricted to using only IoT or edge/cloud computing resources, which can saturate earlier. The effective exploitation of computing resources across the IoT–edge–cloud continuum requires coordinated management of both computing and communication resources, as evidenced by the performance gap between the benchmarks and the *Deterministic* proposal depicted in Figure 1 for both varying system loads and channel conditions.

Figure 1(a) shows that the *Deterministic* proposal can support a larger number of tasks satisfactorily than schemes trying to minimize the execution time of individual tasks. For example, the *Deterministic* proposal can achieve full task satisfaction for up to 60 tasks, whereas the *Minimum* scheme reaches complete satisfaction only up to 45 tasks. The impact of channel conditions reported in Figure 1(b) further highlights the

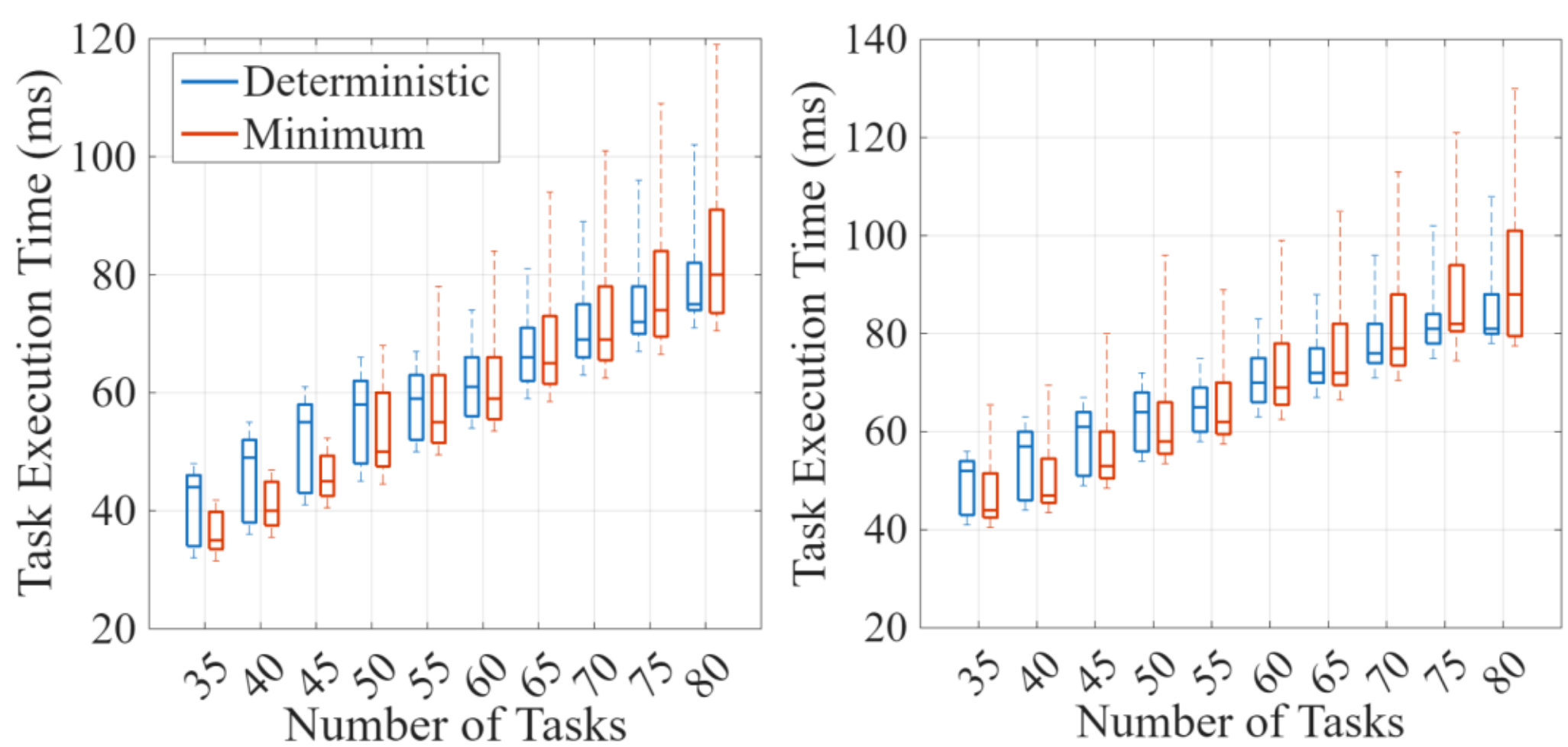

(a) Connected mobility (b) Smart manufacturing

Figure 3. Task execution time distribution for different number of tasks in the connected mobility and smart manufacturing scenarios.

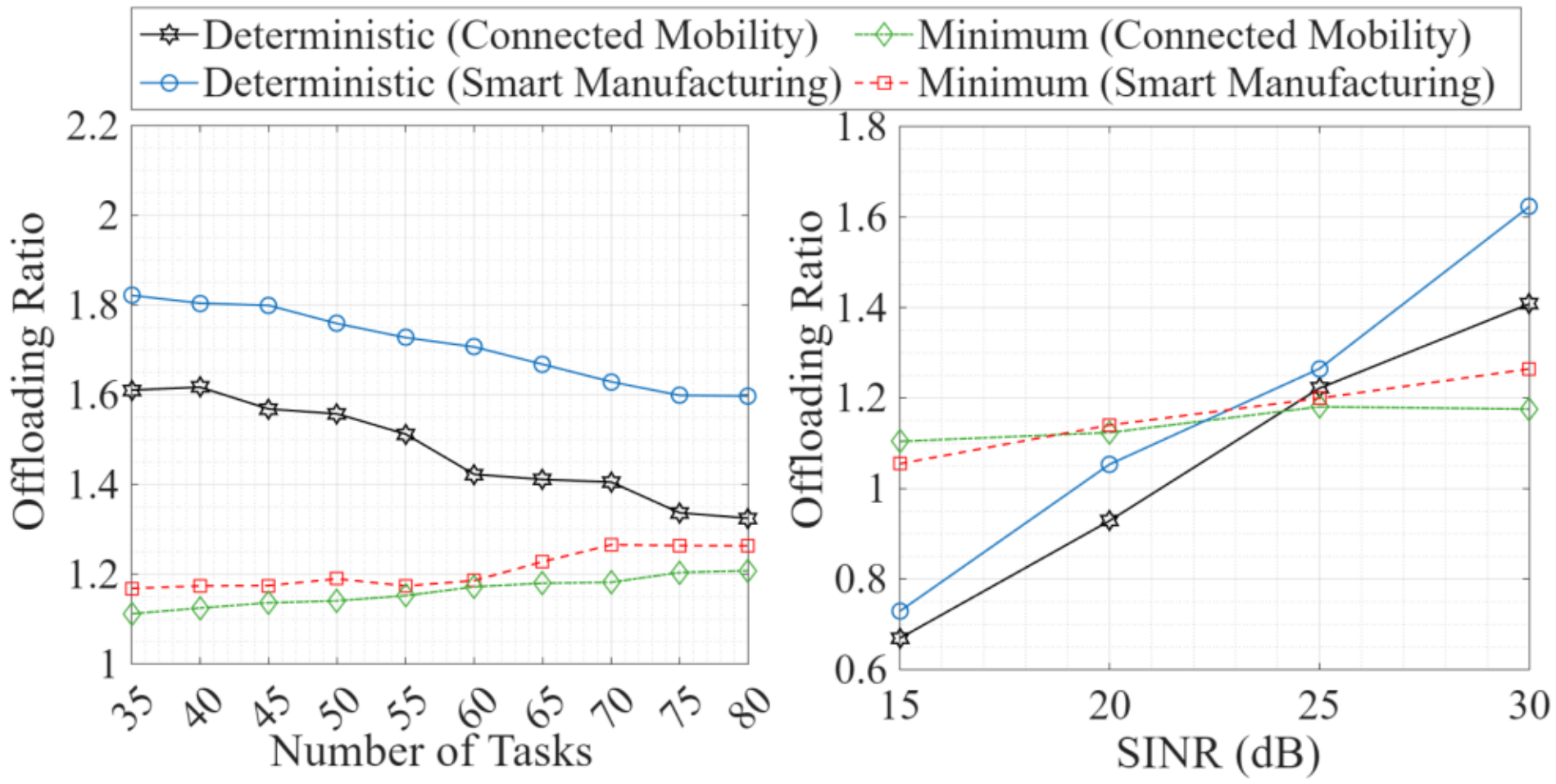

(a) Offloading ratio as a function of the number of tasks (SINR= 30 dB)

(b) Offloading ratio as a function of the SINR (number of tasks=50)

Figure 4. Ratio of offloaded to non-offloaded tasks.

benefits of the proposed *Deterministic* scheme. As the SINR decreases, schemes that rely heavily on offloading suffer from reduced data rates and increased communication congestion. In contrast, the *Deterministic* proposal adapts its task allocation decisions by leveraging local processing when communication resources become constrained. For example, when SINR=15 dB, *Minimum* is offloading 53% of the tasks while the *Deterministic* proposal reduces this percentage to 40%.

The higher satisfaction ratios achieved with the *Deterministic* scheme stem from its better management and more balanced workload and resource utilization across the continuum, which it achieves by flexibly exploiting task deadlines. Figure 2 depicts the average ratio of occupied communication (Figure 2(a)) and computing (Figure 2(b)) resources out of the total resources available between the task arrival time and its deadline. Figure 2 shows that the *Deterministic* proposal balances communication and computing resources usage depending on the workload by spreading tasks more evenly across available resources. In low-load scenarios, the *Minimum* scheme utilizes fewer communication resources and prefers processing more tasks locally compared to the *Deterministic* proposal. However, as the workload increases, the *Minimum* scheme relies excessively on the edge/cloud processing to reduce task's execution time. This

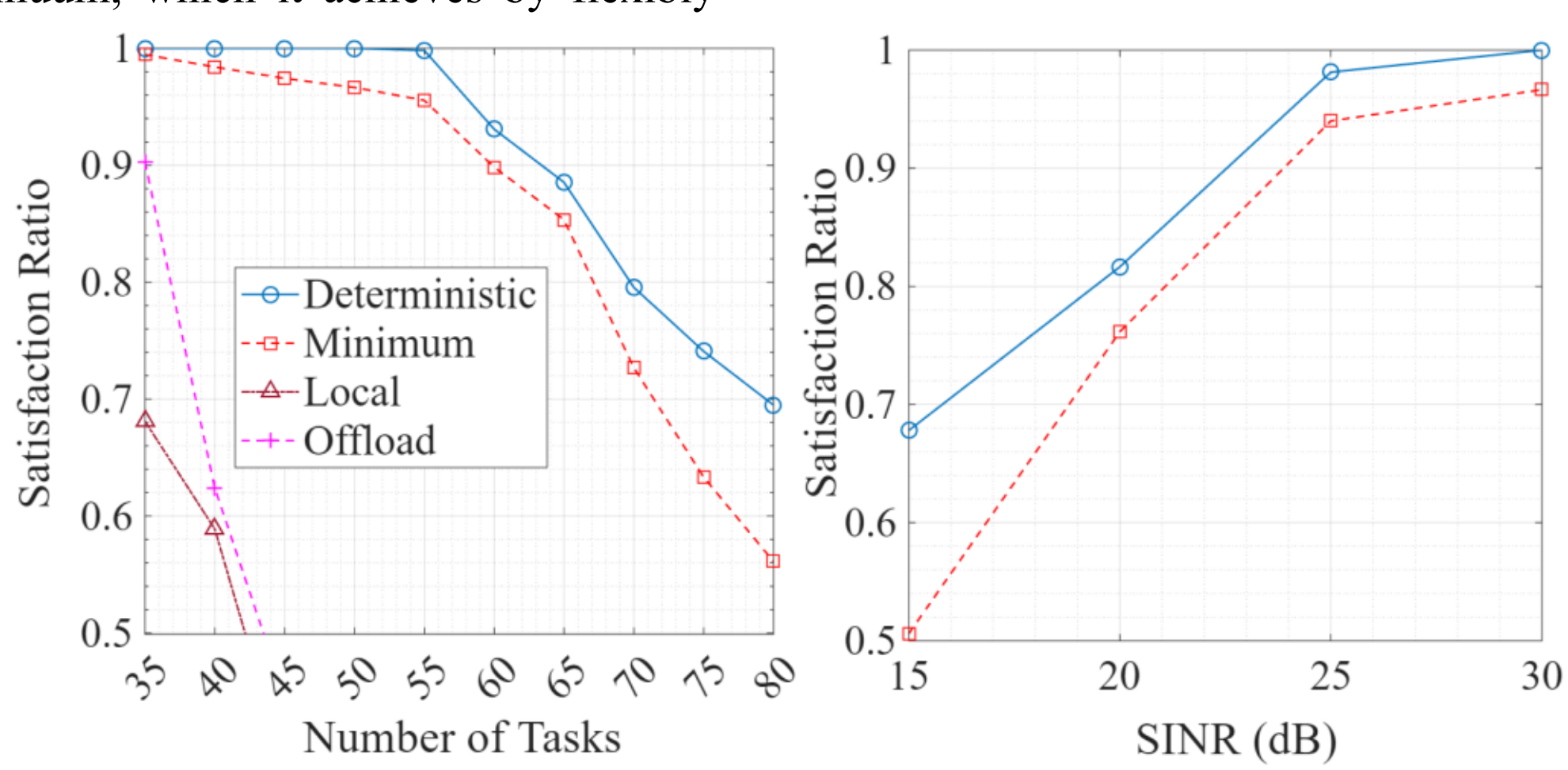

(a) Satisfaction ratio as a function of the number of tasks (SINR= 30 dB)

(b) Satisfaction ratio as a function of the SINR (number of tasks=50)

Figure 5. Satisfaction ratio in the smart manufacturing scenario.

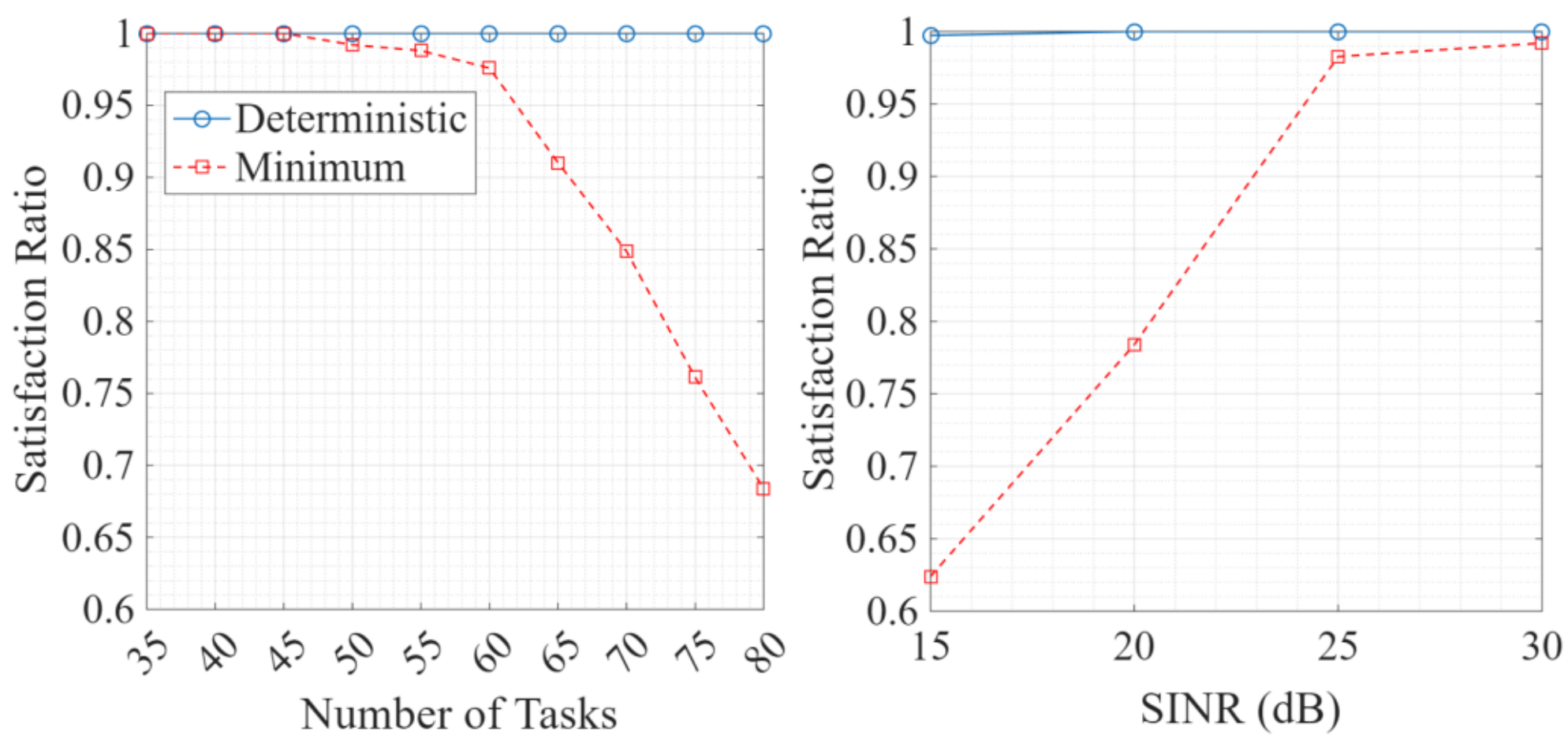

(a) Satisfaction ratio of high-priority tasks as a function of the number of tasks (SINR= 30 dB)

(b) Satisfaction ratio of high-priority tasks as a function of the SINR (number of tasks=50)

Figure 6. Satisfaction ratio of high-priority tasks in the connected mobility scenario.

excessive offloading leads to bottlenecks in the communication links under high loads, resulting in a drop in the satisfaction ratio, as illustrated in Figure 1.

Figure 3 illustrates the task execution time for the *Deterministic* and *Minimum* schemes as a function of the traffic load for both the connected mobility (Figure 3(a)) and smart manufacturing (Figure 3(b)) scenarios. The results are presented as box plots to capture both the median execution time and the variance experienced by the tasks. Specifically, the central mark indicates the median, while the bottom and top edges of the box represent the 25$^{th}$ and 75$^{th}$ percentiles, respectively. The whiskers extend to the minimum and maximum values to capture the variance of task execution times. Under low traffic conditions (e.g., 35 to 45 tasks), the benchmark *Minimum* scheme achieves lower median execution times than the proposed *Deterministic* scheme. This behavior is expected, as the *Minimum* scheme targets the minimization of individual task execution times, exploiting resources when they are available. However, as the number of tasks increases, the performance of the *Minimum* scheme degrades. It leads to resource congestion, resulting not only in higher median execution times but also in an increased execution time variance, as indicated by the elongated upper whiskers in both scenarios at loads of 70 to 80 tasks. In contrast, the proposed *Deterministic* scheme demonstrates superior scalability. By explicitly managing resources to satisfy deadlines rather than minimizing individual execution times, the *Deterministic* scheme prevents severe congestion. Consequently, at high traffic loads, it maintains lower median execution times and a tighter variance compared to the *Minimum* scheme.

Figure 4 compares the operation of the Deterministic and Minimum task offloading and resource allocation schemes under the connected mobility and smart manufacturing scenarios. Both scenarios differ in the computing resources available locally at the IoT devices, which directly influences the balance between local processing and task offloading decisions. Figure 4 depicts the ratio of offloaded to locally processed tasks as a function of the traffic load (Figure 4(a)) and channel conditions (Figure 4(b)). The results show that the Deterministic scheme dynamically adjusts its offloading decisions according to both the workload and available local computing capacity in each scenario. As the number of tasks increases or the SINR decreases, the Deterministic scheme balances the use of local and edge/cloud resources to prevent early resource saturation. In particular, it offloads a higher fraction of tasks to the edge-cloud in the smart manufacturing scenario than in the connected mobility one. This reflects the more constrained local processing capabilities in the smart manufacturing scenario. The difference in offloading ratios between the two scenarios reaches a maximum relative difference of 21% under the highest load condition (i.e. 80 tasks), where computing resources are more stressed. Under these conditions, the Deterministic scheme processes locally a larger proportion of tasks in the connected mobility scenario, leveraging its stronger onboard computing resources. This trend is also visible when analyzing how the offloading ratio evolves with increasing workload. The proportion of offloaded tasks decreases by only 11% in the smart manufacturing scenario while it decreases by more than 21% in the connected mobility scenario. The capacity of the Deterministic scheme to dynamically adjust its allocation decisions results in high task satisfaction ratios for both the connected mobility (Figure 1) and smart manufacturing (Figure 5) scenarios. In contrast, the benchmark Minimum scheme does not explicitly account for differences in local computing capabilities across scenarios in its target to minimize the task execution latency. As a result, it shows largely unchanged offloading behavior despite variations in operating conditions, offloading a similar proportion of tasks in both scenarios (Figure 4). However, since the two scenarios differ in their local processing resources, the Minimum scheme achieves a lower satisfaction ratio in the smart manufacturing scenario than in the connected mobility scenario. The Minimum scheme achieves a 100% satisfaction ratio in the connected mobility scenario for up to 45 tasks, while in the smart manufacturing scenario, it could not achieve 100% even with 35 tasks.

The previous analysis considers scenarios where all tasks have the same priority. Figure 6 compares the satisfaction ratio of high-priority tasks as a function of the number of tasks (Figure 6(a)) and channel conditions (Figure 6(b)) in the connected mobility scenario with equally mixed high- and low-priority tasks. The figure shows that the *Deterministic* proposal can better guarantee deterministic service levels for high-priority tasks under constrained scenarios. In particular, it can satisfy the deadlines of all high-priority tasks, even when there is a high traffic load (i.e., 80 tasks in Figure 6(a)) and constrained communication conditions (SINR=15 dB in Figure 6(b)). On the other hand, trying to minimize the execution time of all tasks reduces the capacity to support deterministic service levels for high-priority tasks. For example, the *Minimum* scheme can only guarantee the execution of 45 high-priority tasks within their bounded deadlines under good communication conditions (SINR= 30 dB). This number reduces by more than 35% when the communication link degrades (SINR=15 dB).

## VII. Conclusions

This study presents a *Deterministic* task offloading and resource allocation scheme for the joint management of communication and computing resources in the IoT–edge–cloud continuum. The proposed scheme prioritizes meeting task deadlines rather than minimizing individual task latency. By leveraging the heterogeneity of task latency requirements, it enables efficient workload distribution and supports a higher number of tasks within their deadlines. In contrast, schemes that aim to minimize the latency of each individual task may fail to efficiently exploit the resources of the IoT–edge–cloud continuum or may impose excessive loads on specific resources. The conducted evaluation demonstrates that a *Deterministic* scheme of task offloading, and resource allocation better adapts to variations in operating conditions, as well as to changes in resource availability and characteristics across the continuum. By dynamically adjusting the balance between local processing and offloading according to scenario-specific capabilities, the *Deterministic* scheme efficiently utilizes available resources across the continuum in diverse operational settings and can guarantee deterministic service levels for high-priority tasks, even under resource-constrained scenarios.